\documentclass[aps,prl,longbibliography,twocolumn,unsortedaddress,superscriptaddress,10pt]{revtex4-1}
\usepackage{graphicx}
\usepackage[normalem]{ulem}
\usepackage{color,bm}
\usepackage{amsmath,amssymb}
\usepackage[T1]{fontenc}
\usepackage{siunitx}
\usepackage{url}

\newcommand{\beq}{\begin{equation}}
\newcommand{\eeq}{\end{equation}}

\usepackage[version=3]{mhchem}

\begin{document}
%Title of paper
\title{3D printed beam splitter for polar neutral molecules}
\author{Sean D. S. Gordon}
\author{Andreas Osterwalder}
\email{andreas.osterwalder@epfl.ch}
\affiliation{Institute for Chemical Sciences and Engineering, Ecole Polytechnique F\'ed\'erale de Lausanne (EPFL), 1015 Lausanne, Switzerland}

\date{\today}

\begin{abstract}
We describe a macroscopic beam splitter for polar neutral molecules. A complex electrode structure is required for the beam splitter which would be very difficult to produce with traditional manufacturing methods. 
Instead, we make use of a nascent manufacturing technique: 3D printing of a plastic piece, followed by electroplating. 
This fabrication method opens a plethora of avenues for research, since 3D printing imposes practically no limitations on possible shapes, and the plating produces chemically robust, conductive construction elements with an almost free choice of surface material; it has the added advantage of dramatically reduced production cost and time.
Our beam splitter is an electrostatic hexapole guide that smoothly transforms into two bent quadrupoles.
We demonstrate the correct functioning of this device by separating a supersonic molecular beam of \ce{ND3} into two correlated fractions. 
It is shown that this device can be used to implement experiments with differential detection wherein one of the fractions serves as a probe and the other as a reference. 
Reverse operation would allow to merging of two beams of neutral polar molecules.
\end{abstract}

\maketitle
\section{Introduction}
Since the first demonstration of the deceleration and trapping of polar neutral molecules\cite{Bethlem:1999ir,Bethlem:2000ti} a large number of experiments have been performed that control the forward velocity and velocity distribution of neutral molecules in a similar manner as with charged particles.\cite{vandeMeerakker:2012ft}
Stark and Zeeman deceleration make use of electric or magnetic field gradients along the direction of propagation of a molecular beam to control the beam velocity.\cite{Bethlem:1999ir, Hudson:2004en, Scharfenberg:2009ht,Vanhaecke:2007ge,Narevicius:2007jw,Osterwalder:2010bx,LavertOfir:2011bf}
Electrostatic quadrupole and hexapole guides have been used in molecular beam experiments for several decades, primarily to filter specific rotational states from within a broader distribution and obtain oriented samples.\cite{Brooks:1976kda,Brouard:2014et}
Recently, such guides have also been used to velocity-filter a molecular ensemble in a curved guide,\cite{Rangwala:2003cy,Junglen:2004eu,Bertsche:2010vr,Bertsche:2011jr} to decelerate a molecular beam in a rotating spiral-guide,\cite{chervenkov:2013vw} and to merge two molecular beams for low-energy scattering studies.\cite{Henson:2012kr,Osterwalder:2015iq,Jankunas:2015eu}

In the present paper we use electrostatic guides to split a molecular beam into two fractions, thus acting only on transverse velocities without changing the longitudinal ones.
In a previous experiment, a microscopic beam splitter on a chip has been demonstrated.\cite{Yin:2007et,Deng:2011ws}
The dimensions of the present, macroscopic, beam splitter match those of most molecular beams experiments and can be used for new types of differential measurement experiments.
For example, one molecular beam can be used as a reference beam while the other is manipulated according to the experimental requirements. 
This can lead to a substantial reduction of measurement times because it avoids the necessity of long data averaging, often needed due to the mechanical properties of the pulsed supersonic valves leading to pulse-to-pulse variations.
In a differential measurement, the effect of these variations is reduced because fluctuations in the parent beam affect probe and reference beams in the same way.
Our device will also be used to superpose two beams by operating it in the reverse configuration to yield precisely the arrangement required for merged-beam studies,\cite{Henson:2012kr,Osterwalder:2015iq,Jankunas:2015eu} which to date have not been possible with two beams of polar molecules due to the difficulty of transversely injecting a beam into an electrostatic guide without the molecules being deflected. 
In a similar manner the present arrangement can be used, for example, to load an electrostatic storage ring for polar molecules\cite{Zieger:2010p5336,Heiner:2009ie} which to date has been possible only by switching the ring off momentarily which imposes odious physical restrictions on the temporal profile of the packet stored inside the ring.

Guides for polar molecules like \ce{ND3} require strong inhomogeneous electric fields which, through the Stark effect, generate a transverse confining force.
In the case of macroscopic guides, the fields are produced by kilovolt-electric potential differences between cylindrical electrodes a few mm apart.
Stringent quality requirements apply to the manufacturing of these electrodes:
a machining precision of better than \SI{50}{\micro\metre} is essential, and the surface finish is critical; impurities, scratches, and similar defects on the electrode surfaces amplify the risk of electric arcing.\cite{Latham:1995tm}
For simple components these requirements can be met by traditional precision machining, but the bent and split shape of the electrodes required here to transform the hexapole guide into quadrupole guides is very challenging to make by traditional  manufacturing techniques.\cite{Chryssolouris:2015et,Wong:2012dm} 
To produce a device that fulfils the mechanical, geometric, and electrical criteria, we apply a modern production method, namely the stereo-lithographic 3D printing of the entire electrode structure as a single plastic piece which then is selectively electroplated with a metal layer $\approx$\SI{10}{\micro\metre} thick. Selective electroplating of the piece allows two electrically independent electrodes to be produced in the correct geometry (see Fig.~\ref{guide}C).
The metal plating not only allows the almost free choice of surface material, including some that would be very hard to machine, it also produces a surface devoid of scratches, recesses or abrasions.

The fabrication method introduced here enables many new avenues for research, beyond gas phase dynamics studies, since 3D printing imposes practically no limitations on possible shapes, and the metal-plating produces chemically robust, conductive construction elements. 
It has the added advantage of dramatically reduced production cost and time: all components used in the present study were printed within less than 48 hours (see video in the supplementary material\cite{thevideo}); electroplating is completed in one day, and the bottle neck of the entire process was the shipping to and from the plating company.
This dramatic acceleration in comparison with traditional manufacturing which, for pieces such as these, can require several months to produce results of lower quality, allows for a very fast turnover and more flexibility in the development and testing of new components.
Furthermore, the entirely digital workflow has the advantage that an inherently exact replica of a complete experimental setup can be produced in any laboratory, simply by transferring a small file, and making use of local production infrastructure.

Several technologies are available for the 3D printing, and the reader is referred to the literature for a full discussion of all the different flavors (\onlinecite{Wong:2012dm} and references therein).
We here use of a technology called stereolithography (SLA).\cite{Bartolo:2011vj}
In SLA a three-dimensional structure is produced in a layer-by-layer process where each layer is painted by a moving laser that photo-polymerizes a precurser inside a resin-filled bath.
The next layer is then added after moving the piece away from the laser by a defined distance and ensuring new resin covers the previous layer.
This process is repeated until the piece is completed.  

Our preference of SLA over alternative technologies is rationalized by the following considerations.
The most commonly known method is fused deposition modelling, where a plastic wire is melted from a fine nozzle that draws the 3D component on a layer-by-layer basis. Extruded molten plastic solidifies and fuses with previous layers upon cooling. Common materials used here generally can be metal plated, but the resolution is inferior to that of SLA and limited by minimum required nozzle diameters and the inherent imprecisions imposed by the melting and fusing of plastic in an atmosphere. 
Selective metal sintering is a technology that produces metallic pieces by selectively melting very fine metal powder in the focus of a laser.\cite{Sames:2016dv} This technology would avoid the requirement of the electroplating, but it is not able to include insulating areas and produces relatively rough surfaces. Mechanical polishing would eliminate one of the principal advantages of our approach.
Several other techniques exist that can provide resolution superior to that of stereolithography, but they use materials that can not be metal coated and are thus not applicable in our experiment, or that are porous and therefore not vacuum-compatible.

\section{Experimental}
\begin{figure}
\includegraphics[width=\columnwidth]{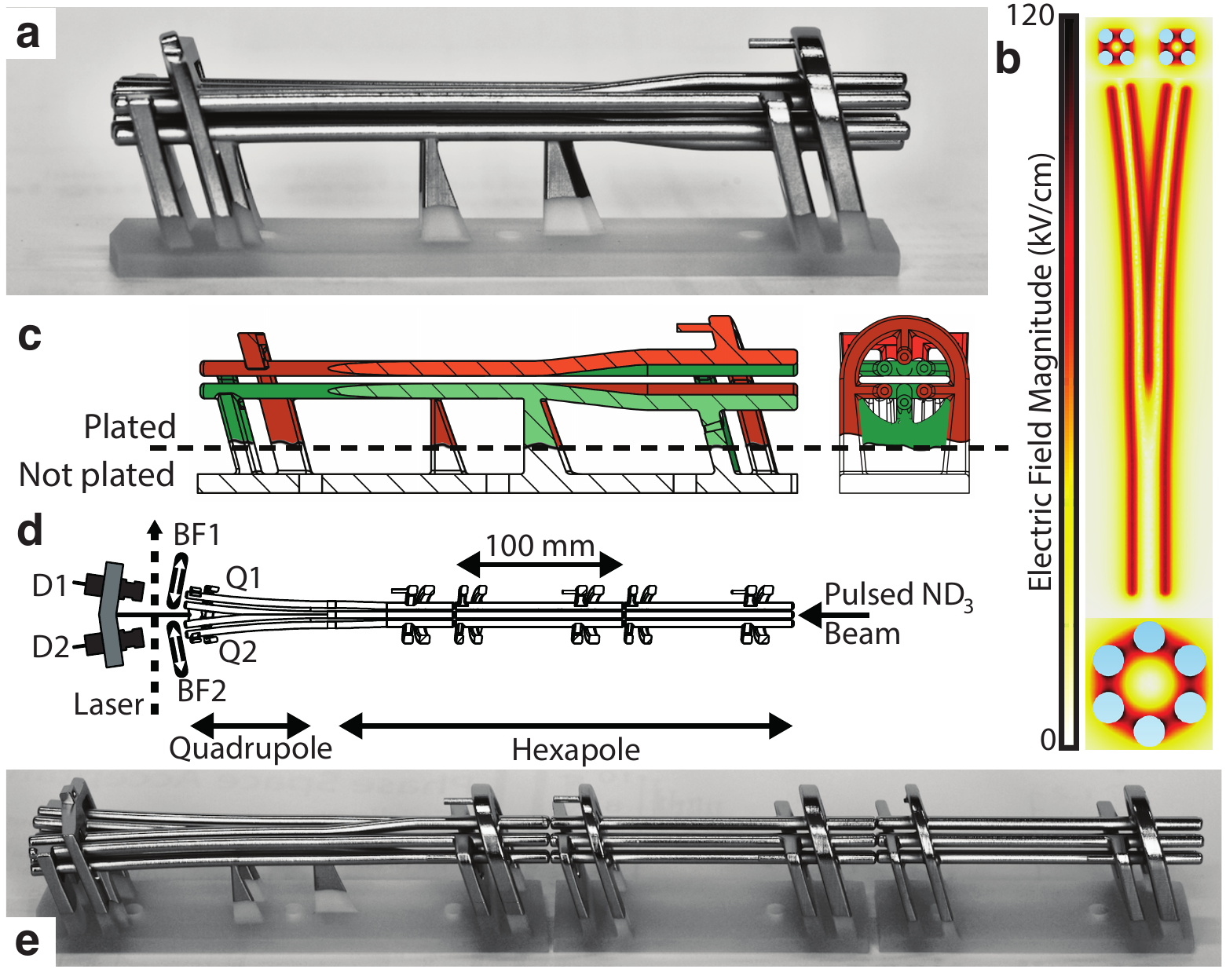}
\caption{\label{guide}a: Side-view of the beam splitter. Molecules are injected from the right. b: Electric field distribution in the guide as a top-view and for cross sections of the hexapole (bottom) and quadrupoles (top), respectively. c: Cross section through the beam splitter. Red and green designate two electrically isolated regions. d: guide and detector arrangement. The \ce{ND3} beam traverses the guide from right to left. A laser beam crosses the molecular beams behind the guide structure to ionise molecules emerging from guides Q1 and Q2. Detectors D1 and D2 record signals from each of the guides individually. Guides can be blocked by retractable beam flags BF1 or BF2. e: Complete guide assembly. }
\end{figure}
The beam splitter itself is shown in figure \ref{guide}a, and the combination of guides used in the present experiment is displayed in \ref{guide}e (standard stereolithography (STL) files of the beam splitter and guides are provided as supplementary material to this paper\cite{suppmatt}).
The top view in figure \ref{guide}d shows the experimental arrangement, housed inside a high-vacuum chamber: a pulsed (General Valve Series 9)  supersonic beam of \ce{ND3} (5\% seeded in Ne, stagnation pressure \SI{2}{\bar}, \SI{20}{\hertz}) is skimmed (diameter \SI{3}{\milli\metre}) and experiences free flight of \SI{235}{\milli\metre} before being fed into the first straight hexapole segment on the right of the figure.
Molecules fly through the hexapole and reach the region where the hexapole  splits into two quadrupoles.
Two separate beams are formed and are detected individually, using two separate channeltron detectors located behind each of the quadrupole guides.
A single focused laser beam crosses both molecular beams and ionizes the \ce{ND3} by rotationally resolved resonance-enhanced multiphoton ionization (REMPI). 
The base pressure obtained in the guide chamber was \SI{8e-8}{\milli\bar}, and rose to \SI{5e-6}{\milli\bar} during operation, sufficient for the present experiments.
While the material used here does not allow for baking as the structural properties break down above $\approx$\SI{80}{\degreeCelsius}, a new resin has become available since these experiments were performed, producing pieces that can be heated to over \SI{250}{\degreeCelsius}.\cite{Anonymous:9Jz9CqDR}
Preliminary tests with this material using plastic components without metal plating permit a base pressure of \SI{6e-10}{\milli\bar} after baking at around \SI{200}{\degreeCelsius} for two days.\cite{Clarke:WPKTzwut}

Different REMPI transitions of \ce{ND3} have been extensively studied in the past.\cite{Ashfold:1998uh,Langford:1998iv,Ashfold:1994us,Allen:1991tb}
The [2+1] scheme chosen here, namely through selected B($\nu_2$) vibronic levels, was used in previous studies in the context of Stark deceleration and velocity filtering of ammonia.\cite{Bertsche:2011jr,Bertsche:2010vr}
In this scheme, two photons around \SI{315}{\nano\metre} excite a transition from the X ground state to a selected B($\nu_2$, J) level which is then ionized by the third photon.
To characterize the guided sample in our experiment we make use of selection rules governing the B$\leftarrow$X transition (as sketched at the top of figure \ref{spectra}): the two components of the inversion doublet in ammonia have differing parity, and the parity of sequential vibrational levels of the $\nu_2$ umbrella mode in the B-state alternates. Consequently, transitions from each of the inversion components can be excited solely to either even or odd vibrational levels. By making use of this technique, we are able to differentiate between \ce{ND3} molecules which emerge from the beam splitter and those which are un-guided and originate directly from the supersonic expansion.

A top-view of the electric field distribution in the main guide element is shown in figure \ref{guide}b.
The single electric field minimum in the hexapole is converted into two field minima in the region of the two quadrupoles, and the \ce{ND3} molecules  are kept on the axes by the edge fields from the guides.
The entire guide structure in figure \ref{guide}a was printed (Formlabs Form 2) as a single poly-methyl methacrylate (PMMA) piece. 
A principal advantage of the 3D printing is that it allows the entire structure to be produced as a single plastic piece. 
This plastic piece is selectively electroplated to produce two separate and electrically isolated electrodes that, during operation, are kept at opposite polarity.
Of the four support structures visible in figure \ref{guide}a and \ref{guide}c to the right and to the left, two (one on either side) are supporting one set of electrodes while the remaining two support the other set of electrodes.
The cone-shaped supports in the middle are connected to a single electrode each.
This enables the separate application of voltages to each of the electrode groups.
Printing the structure as a single piece ensures the relative electrode positions are produced at a precision defined by the resolution of the printer.
The 3D printer used here polymerises a proprietary methyl methacrylate resin and has a specified vertical resolution of \SI{25}{\micro\metre}.
With regards to the horizontal resolution, the supplier specifies the laser spot size to \SI{140}{\micro\metre}, but the printing precision at feature sizes of $>$\SI{200}{\micro\metre} is substantially below \SI{10}{\micro\metre} because the spot size is stable and taken into account during the printing process.\cite{FormlabsInc:vx}

Electric conductivity is obtained from the PMMA piece by selectively coating certain regions with a layer of nickel a few 10s of \si{\micro\metre} thick using a combination of chemical and electrolytical procedures (performed by Galvotec GmbH, Switzerland\cite{galvotec}), thus producing very high quality surfaces.
Since galvanization is a solution-based process, any kind of shadowing is strongly reduced and an even application of the metallic layer is ensured.

\begin{figure}
\includegraphics[width=0.89\columnwidth]{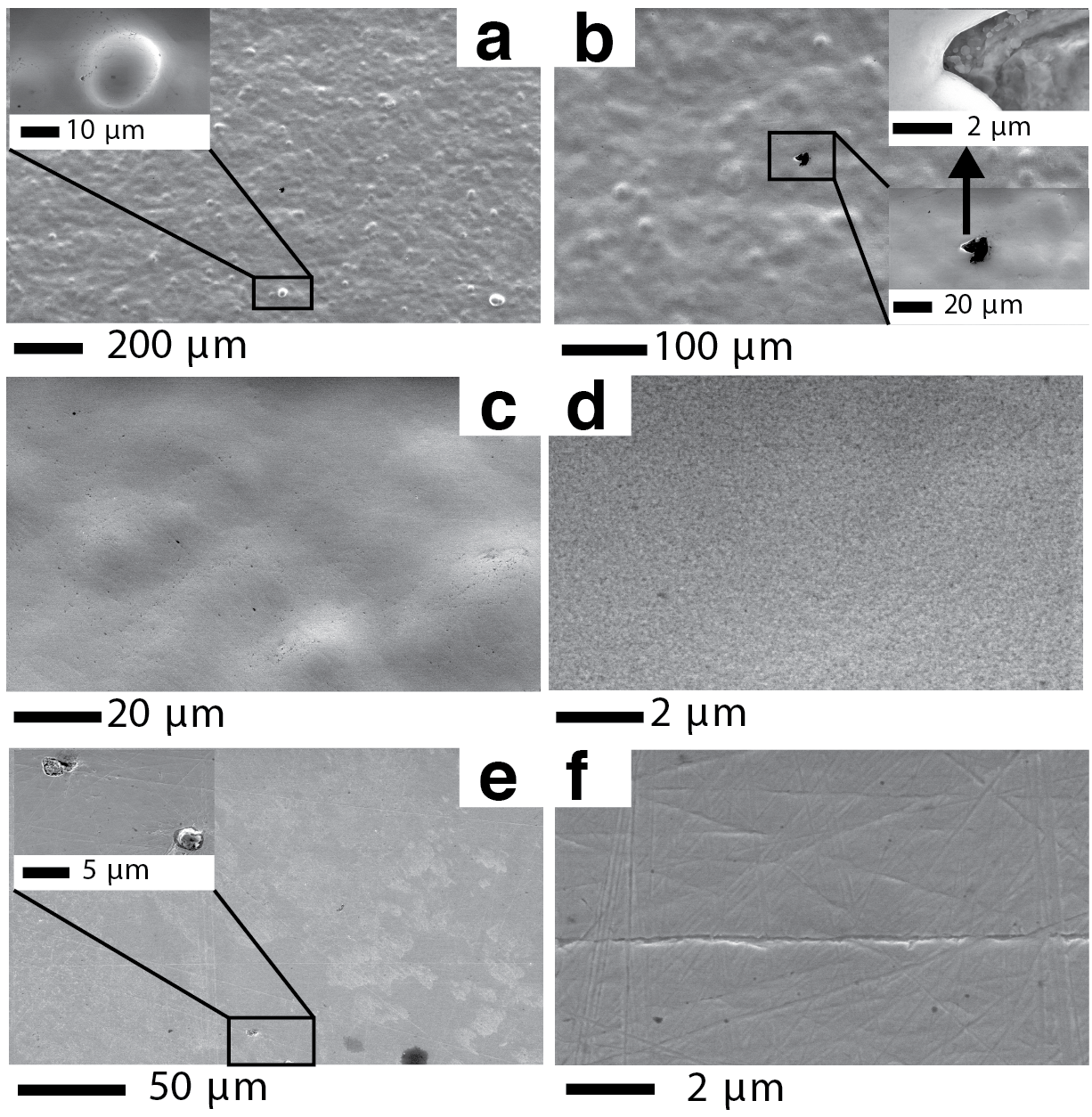}
\caption{\label{micro}Scanning Electron Micrographs of metal plated, 3D-printed PMMA (panels a-d) and of highly polished stainless steel (e and f).}
\end{figure}
Local electric field enhancements at edges or protrusions can lead to extraction of electrons from the bulk metal and thus to arcing.\cite{Latham:1995tm} 
In order to test the high voltage compatibility of the printed parts, we constructed identical test electrodes once from stainless steel and once through printing and galvanisation.
The geometry of the electrodes was such as to mimic a short section of the guides: the closest approach was through rounded, 4 mm diameter edges at a minimum distance of 2 mm. 
The electrodes were  tested with the same power supplies and in the same high vacuum chamber.
Both electrode setups were conditioned up to $\pm$\SI{20}{\kilo\volt} ($\approx$\SI{200}{\kilo\volt\per\centi\metre}), reaching the limit of our power supplies. It is important to note that the conditioning of the printed piece required more time and smaller voltage increments than that of the stainless steel piece. Both electrodes exhibited intermittent arcing during conditioning, but this left no visible trace on either of the electrodes upon subsequent inspection.

Figure \ref{micro} shows electron micrographs of 3D printed, metal-coated pieces (panels a-d) and of polished stainless steel pieces for comparison (panels e and f).
Panels a and b show that the electroplated piece is not perfectly flat and reveals minor  defects,  the major part of the surface is however devoid  of any defects, and in particular, edges and protrusions are absent.
Defects are either  relatively large rounded bumps in the surface, as shown in the inset in panel A, that do not dramatically alter the electric field distribution, or they resemble the cavity shown in the inset of panel b, where the field distortions lead to protected pockets. 
These forms of defects are not believed to efficiently promote electrical breakdown.\cite{Latham:1995tm}
Panels c and d show very smooth surfaces that ensure very good performance of the printed electrodes  in strong electric fields.
In contrast to this, the metal surface in panels e and f shows scratches at a nm scale, and these scratches can ultimately limit the performance of the electrodes in electric fields.
It should be noted that the polishing of these metal pieces was not chosen to produce the highest possible quality but to be representative of the resultant surface quality of complex electrode structures and typical time constraints.
Better surface qualitites undoubtedly are possible, but they require very tedious and slow procedures, and in some cases are limited in terms of structures they can be applied to.

\section{Results}
\begin{figure}
\includegraphics[width=0.9\columnwidth]{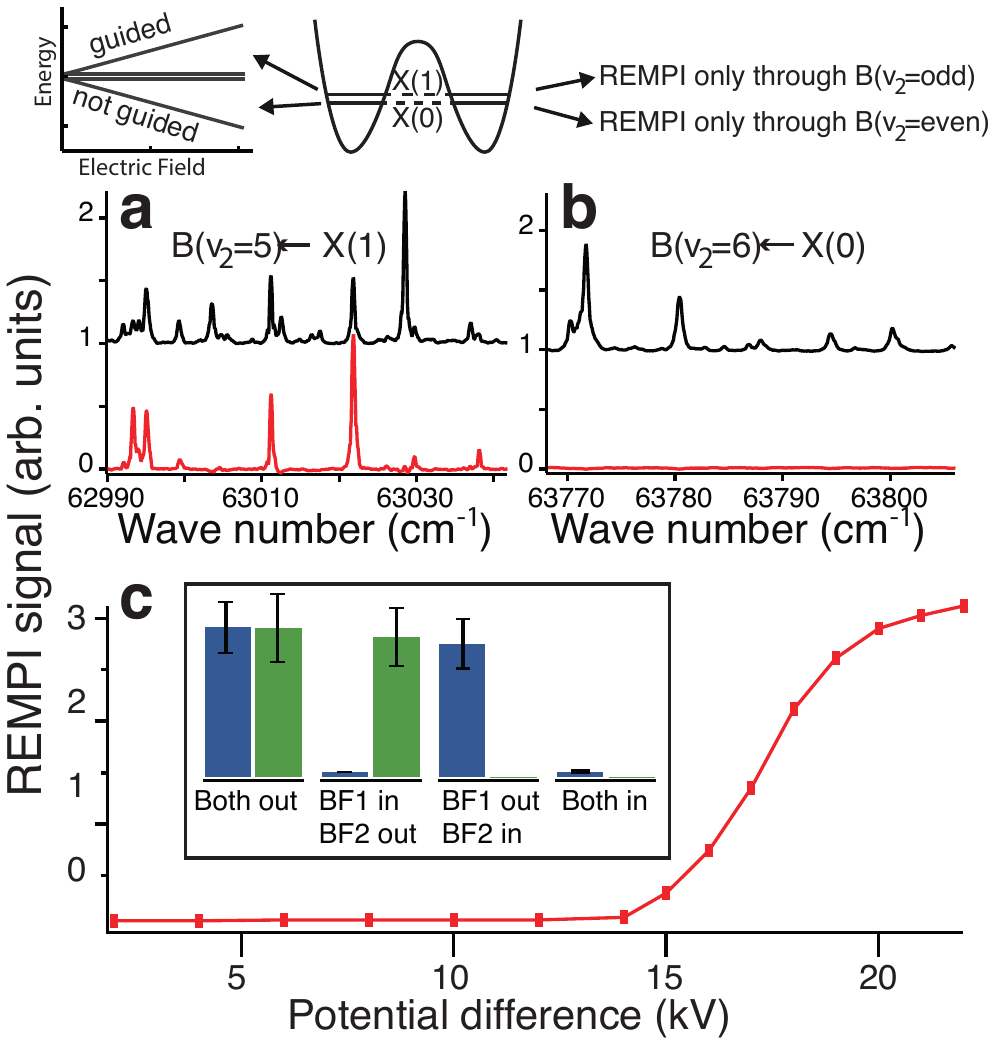}
\caption{\label{spectra}a and b: REMPI spectra recorded through the B($\nu_2$=5) and B($\nu_2$=6) vibrational level, respectively. Black traces: signal of the pure supersonic expansion; red traces: signal from guided molecules. A schematic explanation of the relevant physics is given above the panels (see text). c: Signal as a function of the voltage difference between positive and negative electrodes. Inset: signals on D1 and D2, respectively, when beam flags are as indicated in the figure.}
\end{figure}

\begin{figure}
\includegraphics[width=0.9\columnwidth]{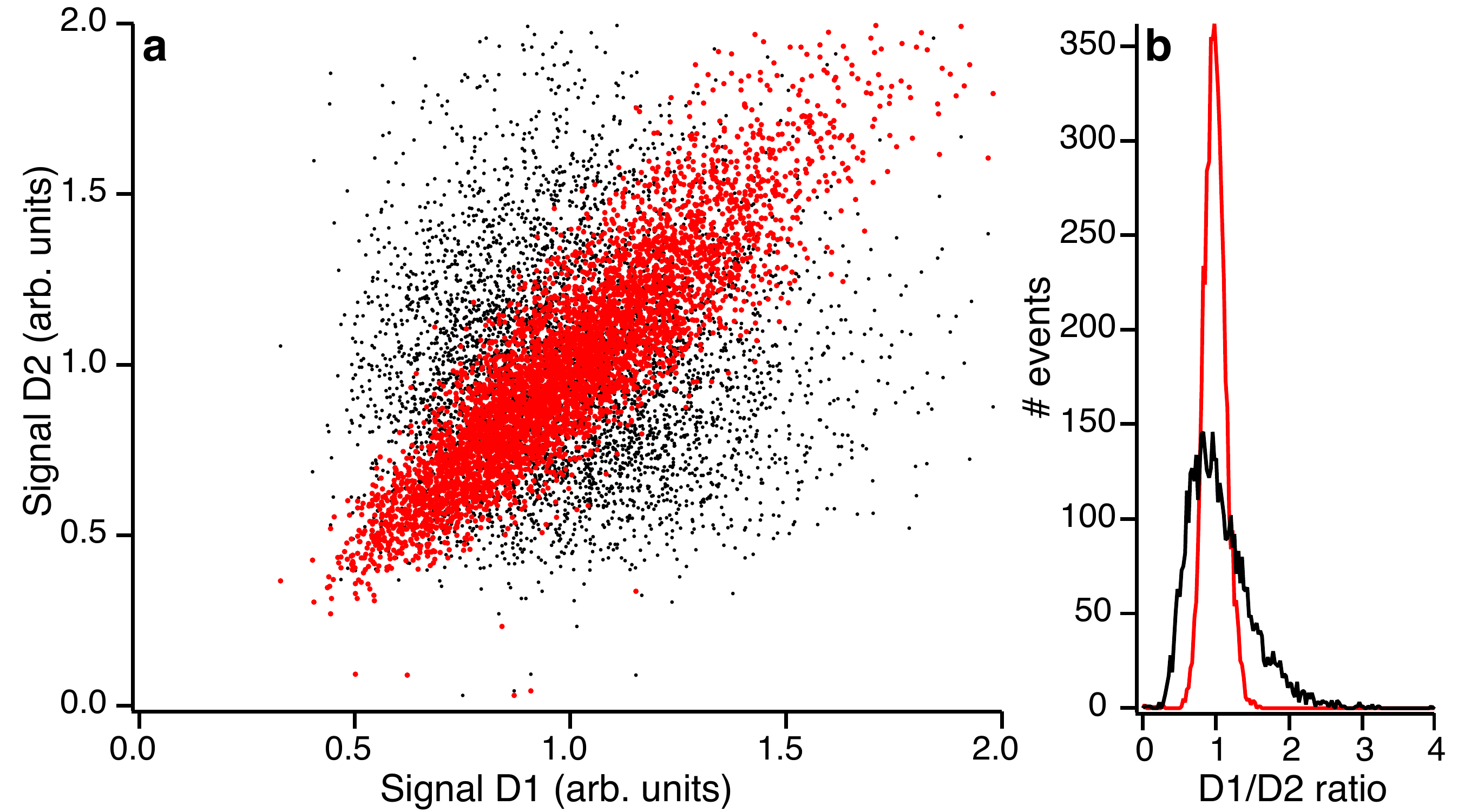}
\caption{\label{results} a: A comparison of the correlation between the normalized signal intensity recorded on detectors D1 and D2 (red) and uncorrelated signal intensity (black). Each point represents one laser shot. b: Histrogram of the ratio between the signals on D1 and D2 for the correlated data (red) and uncorrelated data (black). The correlation coefficient for the correlated data in a is 0.89. The standard deviation of the (un)correlated distribution in b is (40\%)15\%.}
\end{figure}

The principal results of our study are collected in figures \ref{spectra} and \ref{results}.
Panels \ref{spectra}a and \ref{spectra}b show the REMPI signals recorded for the direct supersonic expansion (black traces) and the guided molecules (red traces). At the electric field strength used experimentally, the two levels from the inversion doublet split into Stark states with different projections of the rotational angular momentum vector $\bm J$ on the electric field axis.
Using the quantum number $M_J$ to label these states, we can express the energy of each level as\cite{vandeMeerakker:2012ft}
\begin{equation*}
W_{Stark}=\pm\left(\frac{W_\text{inv}}{2}\mp\sqrt{\left(\frac{W_\text{inv}}{2}\right)^2+\left(\mu_0E\frac{M_JK}{J(J+1)}\right)^2}\right),
\end{equation*}
where $W_\text{inv}$ is the zero-field splitting of the inversion doublet ($W_\text{inv}=\SI{0.053}{\per\centi\metre}$ for \ce{ND3}\cite{Fusina:1985hb}), $\mu_0$ is the permanent electric dipole moment of \ce{ND3} (1.5\,D\cite{DiLonardo:1981dt}), and $E$ is the electric field magnitude.
The force that the states feel is given by the gradient of the potential energy, and points towards stronger fields for levels with $M_JK>0$, and towards weaker electric fields for levels with $M_JK<0$. 
This divides the states into high-field seeking states and low-field seeking states.\cite{vandeMeerakker:2012ft}
Levels with $M_JK=0$ in ammonia have nearly field-independent energies.
Only the upper component of the tunnelling doublet in ammonia, labelled X(1), produces low-field seeking states.
In the electrostatic guides presented here, the electric field is zero in the center and increases towards the edge of the guide, as can be seen in figure~\ref{guide}b.
Low-field seeking states thus feel a force towards the center of the guide and are radially confined, while high-field seeking states are expelled from the center and subsequently ejected from the guide.
Both tunnelling components are essentially equally populated in the original supersonic expansion and the guiding dynamics ensures that only the low-field seeking component is populated after exiting the electric guides.

The REMPI spectrum of the guided sample can no longer contain signal from any transitions through the even $\nu_2$ levels,\cite{Bertsche:2010vr} thus producing a spectroscopic fingerprint of the correct functioning of the beam splitter: as explained above and sketched at the top of figure \ref{spectra}, selection rules allow the two-photon excitation from the X(1) state of B($\nu_2$=odd) levels only, while only B($\nu_2$=even) levels can be excited from the high-field seeking X(0) state.
Several transitions are observed in the black spectra in panels \ref{spectra}a and b, mainly from the J=1, K=0 and 1 rotational levels.
In contrast, only guided states are visible in the red traces, thus simplifying the spectrum of \ref{spectra}a to essentially transitions from the J=1, K=1 state, and reducing the spectrum in \ref{spectra}b to pure background.

To force the \ce{ND3} molecules around the bend in the quadrupole guides, a minimum electric field is required to overcome the centrifugal force.\cite{Junglen:2004eu} In figure \ref{spectra}c the potential difference on the guide electrodes is gradually increased to the maximum of \SI{22}{\kilo\volt} (corresponding to $\approx$\SI{110}{\kilo\volt\per\centi\metre}).
Two principal effects lead to the observed increase in signal: 1. the transverse fields in the entire, straight and curved, guide keep the molecular beam collimated and transport it to the detection region, and 2. in the curved quadrupole section the electric fields compensate for the centrifugal force.
The molecules from the supersonic expansion have a forward velocity of several \SI{100}{\metre\per\second}, and a considerable field is required to bend them around the slight curvature in the current experimental  setup. 
The dynamics leading to the observed signal dependence from the applied voltage is reminiscent of the transmission curves recorded for guide-based velocity filters.\cite{Rangwala:2003cy,Junglen:2004eu,Bertsche:2010vr}
Higher voltages lead to a deeper trap that allows the guiding of faster molecules.
In contrast to the velocity-filter experiments in references \onlinecite{Rangwala:2003cy,Junglen:2004eu,Bertsche:2010vr}, the present velocity distribution is very narrow, and the transmission curve converges when the entire distribution is successfully guided.
The inset in figure \ref{spectra}c shows the signal on detectors D1 and D2 when the beams from Q1 and Q2 are selectively blocked.
By selectively inserting the beam flags in one or both beam paths we confirm that the detected molecules indeed originate from the molecular beam.

In order to perform differential measurements, a correlation is required between probe and reference signals.
In figure \ref{results}a the simultaneously measured, normalised signal intensities from 5000 single-pulse measurements on detectors D1 and D2 are plotted against each other (red dots).
Both signals individually fluctuate, due to shot-to-shot variations of density and detection laser, however there is a clear correlation, and all data points lie near the 45-degree diagonal. 
To compare to the expected result from a single shot measurement, the same data were used to produce an uncorrelated data set by offsetting the D2-signal by one laser shot while keeping that from D1 unaltered (black dots in figure~\ref{results}a). 
As expected, the signal levels are uncorrelated and fill the full plot range of figure~\ref{results}a.
Histograms of the D1/D2 ratios for both the correlated (red) and uncorrelated (black) data sets are shown in figure \ref{results}b. 
The uncorrelated data has a standard deviation of 40\% while the correlated data has a standard deviation of 15\% and a correlation coefficient of $r=0.89$ 
The prime source of the remaining fluctuations in the correlated data is believed to be the detection scheme itself.
The efficiency of a [2+1] REMPI scheme depends non-linearly on the laser power.
While a single laser was focused in the center between the two emerging beams and used for the detection of molecules from both guides, fluctuations in the beam profile, laser power, and position of the focal point would affect the two signals differently.

\section{Summary}
A streamlined development and manufacturing technique making use of 3D printing technology has been used to produce a new type of electrostatic electrode geometry for use in molecular beam experiments. The 3D printing method itself opens a plethora of new possibilities for scientific studies since 3D printing imposes practically no limitations on the shapes that can be produced, and the metal-plating converts any plastic piece into a chemically robust conductive construction element, with the added advantage of reduced cost and production times.
The entirely digital workflow employed here furthermore offers the possibility to produce an inherently exact replica of a complete experimental setup in any laboratory, simply by transferring a small file and employing local production infrastructure.
This infrastructure is a moderately priced piece of equipment that requires no specialized personnel and can thus also be operated at places that do not offer sophisticated workshop infrastructure.

We demonstrate that our device allows to cleanly split a molecular beam into two fractions of comparable and correlated densities and thus is is capable of reducing shot-to-shot signal fluctuation. 
These measurements point towards the possibility of using the device for probe and reference type experiments, in which one of the beams is used experimentally and the other beam serves as a blank.  
Rotating the beam splitter by 180 degrees and injecting two separate supersonic expansions will enable the merging of two beams of polar molecules and allow low-temperature reactivity studies between them.
Merged beam experiments allow reactions to be studied at collision energies substantially below \SI{1}{\kelvin}.\cite{Henson:2012kr,Osterwalder:2015iq,Jankunas:2014hg}
The current merged beam experiments superpose a magnetically guided beam with either an electrostatically guided or a non-guided secondary beam.
Because the electric fields from a guide are present both inside and outside the electrode structures, the analogous injection of a polar beam into an electrostatic guide would be very difficult.
In contast, merging a beam using the inverted beam splitter will be a much less challenging experiment to perform, and the detailed investigation of dipole-dipole interactions in merged molecular beams is now within reach.

\section{Acknowledgments}
We thank Mr. Rico Schuhmacher (Galvotec, Switzerland) for his expert help in electroplating the 3D printed pieces, Ms. Christiane Fimpel (3D-Model, Switzerland) for her support in our initial testing of 3D printing, and Dr. Gregoire Baroz from the EPFL Interdisciplinary Centre for Electron Microscopy (CIME) for his expert assistance.
This work is funded by EPFL and the Swiss Science Foundation (project number 200021\_165975).

%\bibliographystyle{aipnum4-1}
%

%\bibliography{../../Allrefs.bib,../../Allrefs2.bib}

\end{document}